\documentstyle[12pt,aasms4]{article}

\begin{document}

\title{Precise Interplanetary Network Localization of a New Soft
Gamma Repeater, SGR1627-41}

\author{K. Hurley}
\affil{University of California, Berkeley, Space Sciences Laboratory,
Berkeley, CA 94720-7450}
\authoremail{khurley@sunspot.ssl.berkeley.edu}
\author{C. Kouveliotou} 
\affil{Universities Space Research Association at NASA Marshall Space Flight Center, 
ES-84, Huntsville AL 35812}
\author{P. Woods\altaffilmark{1}}
\affil{NASA Marshall Space Flight Center, ES-84, Huntsville, AL 35812}
\author{E. Mazets, S. Golenetskii, D. D. Frederiks}
\affil{Ioffe Physical-Technical Institute, St. Petersburg, 194021, Russia}
\author{T. Cline}
\affil{NASA Goddard Space Flight Center,
Greenbelt, MD 20771}
\author{J. van Paradijs\altaffilmark{2}, University of Alabama in Huntsville, AL 35899}
\altaffiltext{1}{University of Alabama in Huntsville, AL 35899}
\altaffiltext{2}{Astronomical Institute `Anton Pannekoek',
University of Amsterdam, The Netherlands}

\begin{abstract}

We present \it Ulysses \rm, KONUS-WIND, and BATSE observations of bursts from a new soft gamma
repeater which was active in 1998 June and July.  Triangulation of the bursts results
in a $\sim$ 1.8 $\arcdeg$ by 16 $\arcsec$ error box whose area is $\sim$ 7.6 arcminutes$^2$,
which contains the Galactic supernova remnant G337.0-0.1.  This error box intersects the
position of a BeppoSAX X-ray source which is also consistent with the position of
G337.0-0.1 (Woods et al. 1999), and is thought to be the quiescent
counterpart to the repeater.  If so, the resulting error box is $\rm \sim 2 \arcmin \times
16 \arcsec$ and has an area of $\sim$ 0.6 arcminutes$^2$.  The error box location within
the supernova remnant suggests that the neutron star has a transverse velocity of
$\sim$200 - 2000 km/s.

\end{abstract}


\section{Introduction}

There is good evidence that the three known soft gamma repeaters (SGRs) are associated
with supernova remnants (SNRs).  SGR0525-66 appears to be in 
N49 in the Large Magellanic Cloud (Cline et al. 1982), and SGR1806-20 (Atteia et al. 1987) in  
G10.0-0.3 (Kulkarni \& Frail 1993; Kouveliotou et al. 1994; Kulkarni et al. 
1994; Murakami et al. 1994).    SGR1900+14 lies
close to, although not within, G42.8+0.6 (Kouveliotou et al. 1994; Hurley et al. 1999a).
The SGRs are believed to be magnetars, i.e. neutron stars with magnetic field strengths
in excess of 10$^{14}$ G.  In these objects, magnetic energy dominates rotational energy.  

In this paper we present gamma-ray observations of a new source, SGR1627-41, first detected in
1998 June, 
whose repetition, time histories, energy spectra, and location are all consistent with
the properties of the known SGRs.  We present observations by the 3rd interplanetary
network (IPN), consisting in this case of the gamma-ray burst
experiment aboard the \it Ulysses \rm spacecraft, the KONUS experiment aboard the
WIND spacecraft, and the \it Burst and Transient Source Experiment \rm (BATSE) aboard
the \it Compton Gamma-Ray Observatory \rm (CGRO), and use them to derive a precise
source location for SGR1627-41.  The typical energy ranges for the experiments and the data considered here are
25-150, 50-200, and 25-100 keV, respectively.  At the time of these
observations, \it Ulysses \rm was located $\sim$ 2900 light-seconds from earth while
WIND was $\sim$ 3 light-seconds away; CGRO was in near-Earth orbit.

\section{Observations}

The first confirmed \it Ulysses \rm observation of the new SGR was on
1998 June 17.  The last was on 1998 July 12.  In all, 36 events were
observed by \it Ulysses \rm and an instrument on at least one near-earth spacecraft, either
KONUS-WIND or BATSE, and triangulation gave mutually consistent annuli.  (Numerous other
events were observed by BATSE alone, and numerous candidate events were also observed by a single instrument, either KONUS or \it Ulysses \rm GRB, which could
not be localized; we do not consider them here.)  Each of the three instruments
has various data-collecting modes which may be summarized as either ``triggered'' or
``untriggered''.  The time resolutions in triggered modes are as fine as 2 ms, while
in untriggered modes, they may be as coarse as $\sim$ 1 s.  Table 1 lists the bursts and
the modes, and figure 1 shows a typical time history.  
When observed in triggered mode by \it Ulysses \rm, almost all of the events considered here had durations $\lesssim$ 200 ms; as observed by BATSE, several events had durations
of up to $\sim$ 2 s.

In principle, short events such as these present an ideal case for localization
by triangulation, since the width of a triangulation annulus is proportional to
the uncertainty in cross-correlating the time histories observed by a pair of
spacecraft.  However, two other factors must be considered.  First, to obtain
a small cross-correlation uncertainty, the event must be observed in triggered
modes by \it Ulysses \rm and another spacecraft; only 11 of the events in table
1 satisfy this criterion.  (Intense activity from a repeating source tends to
fill trigger memories, so that subsequent events can only be recorded in
an untriggered mode.)  Second, the proximity of the WIND and CGRO spacecraft
results in triangulation annuli which intersect at grazing incidence for any
given burst, resulting in a long, narrow error box.  To reduce the length of
the error box for this SGR, we have combined bursts from the first and last
triggered mode observations, on June 17 and July 12.  The slowly moving \it Ulysses \rm-
Earth vector, which is approximately the center of the triangulation
annulus, produces a shorter error box.

\section{Results}

Figure 2 shows a portion of the IPN error box, defined by the intersection of
two $\sim$ 16 $\arcsec$ wide annuli.  The corners of this $\sim$ 7.6 arcminute$^2$
error box are given in table 2.  Strictly speaking, the curvature of the annuli
does not allow the resulting error box to be defined by straight-line segments;
for this reason, we also give the centers, radii, and widths of the annuli in
table 3.  Figure 2 also includes the 843 MHz radio contours of the Galactic supernova remnant
G337.0-0.1, taken from the catalog of Whiteoak and Green (1996).  Finally,
figure 2 also shows the position of a BeppoSAX quiescent X-ray source believed
to be the SGR counterpart (Woods et al. 1999).  The intersection of the 3$\sigma$ IPN
annuli with the 95\% confidence error circle defines a $\rm \sim 2 \arcmin \times
16 \arcsec$ error box whose area is $\sim$ 0.6 arcminutes$^2$; the coordinates of
this error box are given in table 4.

These two error boxes are consistent with, but considerably
smaller than, the following locations previously determined for this SGR:
1) the BATSE error circle derived from four triggers (Kouveliotou et al. 1998a). 
(Based on this initial location, the source was named SGR 1627-41).
2) the initial IPN annulus (Hurley et al. 1998a)
3) the restriction of the initial IPN annulus to locations
consistent with BATSE earth-limb occultation considerations (Woods et al. 1998)
4) a refined, but still preliminary IPN annulus (Hurley et al. 1998b), 
5) the initial \it Rossi X-Ray Timing Explorer All Sky Monitor (RXTE-ASM) \rm error box (Smith and
Levine 1998), and
6) the final RXTE-ASM error box (Smith et al. 1999)

\section{Discussion}

If we adopt as a working hypothesis that SGR1627-41 is a magnetar, and
that magnetars have lifetimes $\sim$ 10,000 years (Thompson
and Duncan 1995), then we would expect the IPN position to be coincident
with that of a radio supernova remnant, whose observable lifetimes
are $\lesssim$ 20,000 years (Braun, Goss, and Lyne 1989).  Also,
three SGRs are known to be quiescent
soft X-ray point sources: SGR0525-66 (Rothschild, Kulkarni, \& Lingenfelter 1994),
SGR1806-20 (Murakami et al. 1994), and SGR1900+14 (Hurley et al. 1994, 1999b). 
With this in mind, we
can then inquire how compelling the IPN/G337.0-0.1/BeppoSAX association is.
We first calculate the probability  
that the 1.8$\arcdeg$ by 16$\arcsec$ IPN annulus intersects a SNR in the 843 MHz survey of the
MOST catalog (Whiteoak and Green 1996)?  A rigorously correct method
to estimate this probability was presented by Kulkarni and Frail (1993),
but a very simple argument can be used to derive an absolute lower limit.  
The MOST survey covered galactic
coordinates $\rm 245\arcdeg < l < 355\arcdeg, \mid b \mid < 1.5\arcdeg$.
73 SNRs with measured sizes are cataloged, and they occupy $\sim$ 0.022
of the total area surveyed.  Thus in the limit where the error box is
a point, the probability of a chance association
would be $\sim$ 0.022.  However, two factors will increase this substantially.
First, given the fact that SGR1900+14 appears to be outside its
supernova remnant (Hurley et al. 1999a), we would probably accept an SGR/SNR
association where the error box lay outside the remnant, increasing the
effective occupied area of the survey.  Second, the method of Kulkarni and Frail (1993),
which is more appropriate for the long, narrow IPN error box, would result 
in a higher probability.

We next ask what the probability is that the IPN error box
will coincide with a quiescent soft X-ray source.  One unidentified source 
with a 1 $\arcmin$ error radius was 
detected in the BeppoSAX observations (Woods et al. 1999), and the
field of view was 28 $\arcmin$ in radius.    Applying the method of Kulkarni and
Frail (1993) gives a chance probability of 0.17.  Thus the joint probability of
the IPN/G337.0-0.1/BeppoSAX association is $>$0.004.

Fortunately, more data which can substantiate the IPN/X-ray source/SNR association
are forthcoming.  An \it Advanced Satellite for Cosmology and 
Astrophysics \rm observation of the X-ray source is planned for 1999.
This will allow us to confirm the suggested 6.47 s period, observed with
a low statistical significance in the BeppoSAX data (Woods et al. 1999),
and possibly derive the period derivative.  A high spindown
rate, as found for SGR1806-20 and SGR1900+14 (Kouveliotou et al. 1998b, 1999) would
be a compelling argument that the source is indeed a magnetar associated with
the SNR.

If this indeed proves to be the case, the transverse velocity of
the magnetar can be estimated.  The distance to G337.0-0.1 has been
estimated to be as small as 5.8 kpc by Case and Bhattacharya (1998) 
based on a
a new $\Sigma$-D relation for supernova remnants, and as large as
11 kpc by Sarma et al. (1997) based on radio recombination lines.
The displacement between
the core of the remnant and the IPN/SAX error box is $\sim$1.3 $\arcmin$.
From this, we obtain velocities between $\sim$200 and 2000 km/s for the
smaller distance, and for assumed
ages of 10,000 and 1000 y, consistent with the transverse velocities of
the other three SGRs.

\acknowledgments
KH is grateful to JPL for Ulysses support under Contract 958056,
and to NASA for Compton Gamma-Ray Observatory support under
contract NAG 5-3811.

\newpage

\figcaption{Time history of the burst on 1998 June 18 04:30:29, from the Ulysses
GRB experiment.  The energy range is 25-150 keV. \label{fig1}}

\figcaption{Two IPN annuli superimposed on the 843 MHz radio contours of
G337.0-0.1 from Whiteoak and Green (1996).  They characterize the
remnant as a peculiar non-thermal object, part of which may be
extragalactic.  The BeppoSAX error circle is also shown (Woods et al. 1999).
Given the rough alignment of the two radio lobes and
the IPN/BeppoSAX error box, we speculate that another explanation of
the morphology might be an asymmetric
supernova explosion. \label{fig2}
}

\clearpage
\begin{deluxetable}{ccccc}
\tablecaption{\it Observations of SGR1627-41.  }
%\tablewidth{0pt}
\tablehead{
\colhead{Date} & \colhead{UT\tablenotemark{1}}  & \colhead{\it Ulysses \rm} & \colhead{BATSE} & \colhead{KONUS}
}
\startdata
17 JUN 98 &	18:53:25 & T\tablenotemark{2}	& 6832\tablenotemark{3}	& U	\nl
17 JUN 98	&	18:58:14	&	U\tablenotemark{4} & U	&	U	\nl
17 JUN 98	&	19:08:12	&	U	&	U	&	NO	\nl
17 JUN 98	&	19:58:31	&	T	&	O\tablenotemark{5}	&	T	\nl
17 JUN 98	&	20:06:30	&	U	&	U	&	U	\nl
17 JUN 98	&	20:17:47	&	U	&	SAA\tablenotemark{6}	&	U	\nl
17 JUN 98	&	20:57:26	&	T	&	6833	&	NO	\nl
17 JUN 98	&	21:04:40	&	U	&	U	&	T	\nl
17 JUN 98	&	21:07:24	&	U	&	O	&	T	\nl
17 JUN 98	&	21:17:06	&	U	&	O	&	U	\nl
17 JUN 98	&	21:37:18	&	U	&	6834	&	U	\nl
17 JUN 98	&	21:57:10	&	T	&	SAA	&	U	\nl
17 JUN 98	&	22:05:57	&	U	&	SAA	&	U	\nl
17 JUN 98	&	22:54:08	&	U	&	O	&	T	\nl
18 JUN 98	&	00:15:35	&	U	&	O	&	U	\nl
18 JUN 98	&	01:42:31	&	U	&	6837	&	T	\nl
18 JUN 98	&	03:07:49	&	T	&	U	&	U	\nl
18 JUN 98	&	03:28:30	&	U	&	O	&	T	\nl
18 JUN 98	&	03:53:51	&	U	&	6838	&	NO	\nl
18 JUN 98	&	03:54:13	&	U	&	U	&	NO	\nl
18 JUN 98	&	03:59:39	&	U	&	U	&	U	\nl
18 JUN 98	&	03:59:55	&	U	&	U	&	NO	\nl
18 JUN 98	&	04:02:04	&	U	&	U	&	NO	\nl
18 JUN 98	&	04:04:21	&	U	&	6839	&	U	\nl
18 JUN 98	&	04:30:27	&	U	&	U	&	T	\nl
18 JUN 98	&	04:30:29	&	T	&	U	&	T	\nl
18 JUN 98	&	04:34:20	&	U	&	U	&	U	\nl
18 JUN 98	&	06:16:33	&	T	&	U	&	T	\nl
18 JUN 98	&	07:34:06	&	T	&	6841	&	T	\nl
18 JUN 98	&	16:38:04	&	T	&	SAA	&	T	\nl
18 JUN 98	&	17:51:16	&	T	&	O	&	U	\nl
22 JUN 98	&	13:29:56	&	T	&	O	&	T	\nl
22 JUN 98	&	14:11:24	&	U	&	6861	&	NO	\nl
22 JUN 98	&	18:56:37	&	T	&	6862	&	T	\nl
25 JUN 98	&	10:56:19	&	T	&	SAA	&	T	\nl
12 JUL 98	&	21:50:39	&	T	&	6919	&	T	\nl

\enddata
\tablenotetext{1}{Time at GRO or WIND}
\tablenotetext{2}{Recorded in triggered mode}
\tablenotetext{3}{BATSE trigger number}
\tablenotetext{4}{Recorded in untriggered mode}
\tablenotetext{5}{The source was occulted by the Earth}
\tablenotetext{6}{The spacecraft was passing through the South Atlantic Anomaly
(SAA) and the high voltage was turned off}
\end{deluxetable}

\clearpage
\begin{table*}
\begin{center}
\begin{tabular}{cc}
$\rm \alpha(2000)$, degrees& $\rm \delta(2000), degrees$ \\
\tableline
248.9621 & -47.6015 \\
248.9503 & -48.4118 \\
248.9891 & -46.6026 \\
248.9707 & -47.5451 \\
\end{tabular}
\tablenum{2}
\caption{Corners of the IPN error box for SGR1627-41}
\end{center}
\end{table*}

\clearpage
\begin{deluxetable}{ccccc}
\tablecaption{Parameters of the two IPN annuli defining the error box}
\tablehead{
\colhead{Date} & {$\rm \alpha(2000)_{center}$, \arcdeg}& \colhead{$\rm \delta(2000)_{center} \arcdeg$} & \colhead{Radius$\arcdeg$} & \colhead{3$\sigma$ half-width \arcdeg}
}
\startdata
1998 Jun 17 & 330.4126 & -10.1705 & 76.7554 & 0.002599 \\
1998 Jul 12 & 332.6195 & -8.3063 & 79.6049  & 0.002440 \\
\enddata
\end{deluxetable}

\clearpage
\begin{table*}
\begin{center}
\begin{tabular}{cc}
$\rm \alpha(2000)$, degrees& $\rm \delta(2000), degrees$ \\
\tableline
248.9618 & -47.6121 \\
248.9626 & -47.5792 \\
248.9691 & -47.6106 \\
248.9698 & -47.5808 \\
\end{tabular}
\tablenum{4}
\caption{Corners of the IPN/BeppoSAX error box for SGR1627-41}
\end{center}
\end{table*}

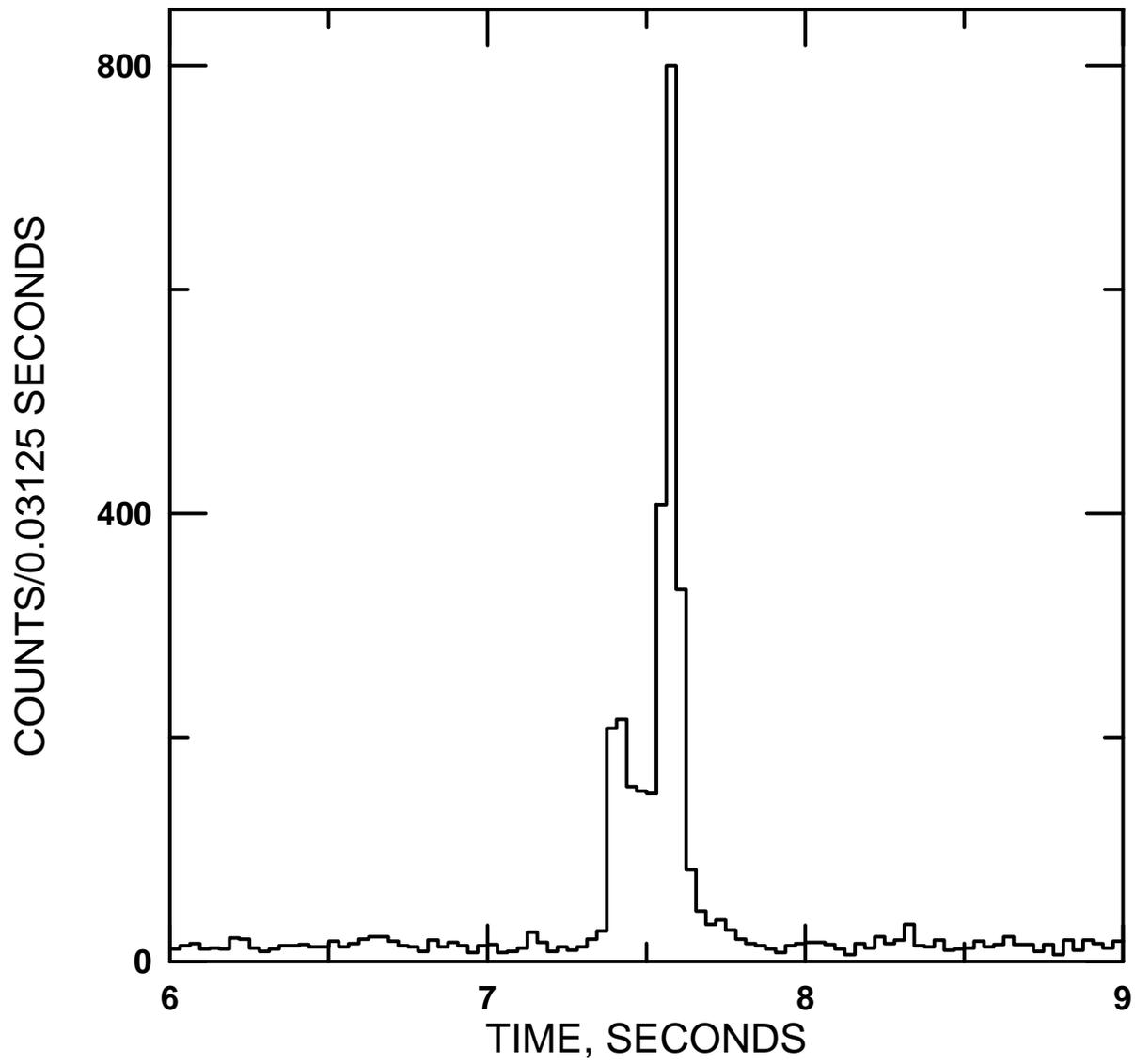
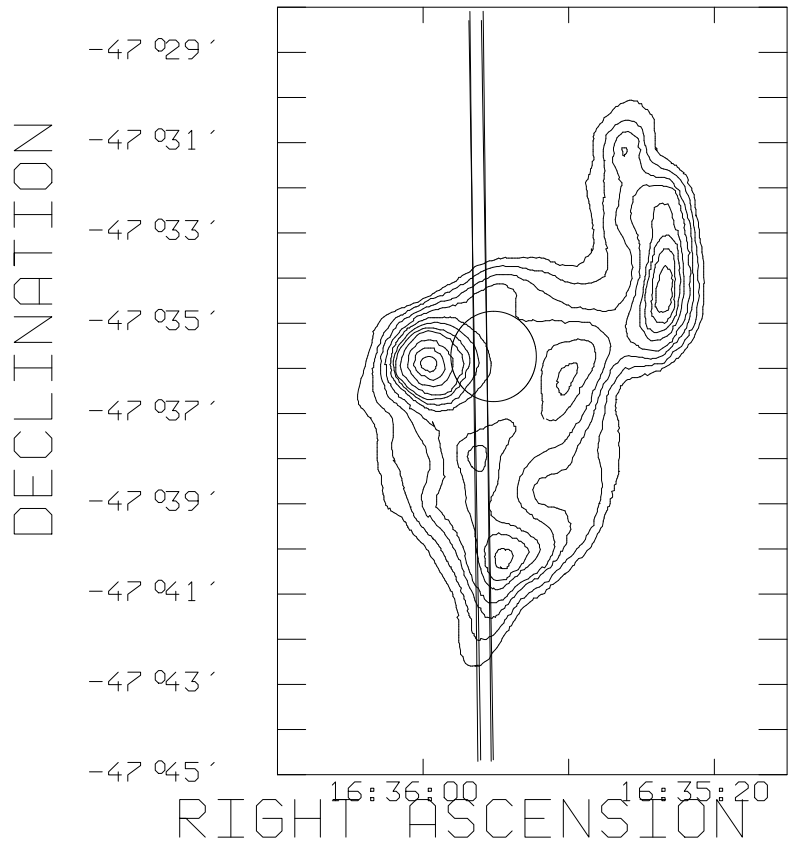
\end{document}